Nonlinear quantum-mechanical system associated with Sine-Gordon equation in (1+2) dimensions

Yair Zarmi

Jacob Blaustein Institutes for Desert Research
Ben-Gurion University of the Negev
Midreshet Ben-Gurion, 84991 Israel

Abstract

Despite the fact that it is not integrable, the (1+2)-dimensional Sine-Gordon equation has $N$-soliton solutions, whose velocities are lower than the speed of light ($c$ =1), for all $N \geq 1$. Based on these solutions, a quantum-mechanical system is constructed over a Fock space of particles. The coordinate of each particle is an angle around the unit circle. U, a nonlinear functional of the particle number-operators, which obeys the Sine-Gordon equation in (1+2) dimensions, is constructed. Its eigenvalues on $N$-particle states in the Fock space are the slower-than-light, $N$-soliton solutions of the equation. A projection operator (a nonlinear functional of U), which vanishes on the single-particle subspace, is a mass-density generator. Its eigenvalues on multi-particle states play the role of the mass density of structures that emulate free, spatially extended, relativistic particles. The simplicity of the quantum-mechanical system allows for the incorporation of perturbations with particle interactions, which have the capacity to "annihilate" and "create" solitons - an effect that does not have a classical analog.

Email: zarmi@bgu.ac.il

## 1. Introduction

The Sine-Gordon equation has attracted extensive attention for decades in the study of classical and quantum-mechanical phenomena [1-9]. In parallel, it has played an important role within the framework of Quantum-Field Theory [9-20], where quantization of the classical equation was a major issue, the goal being the mapping of the classical solution onto a field operator.

In this paper, a different procedure for the construction of a quantum-dynamical system that is associated with the Sine-Gordon equation in (1+2) dimensions is presented. The procedure does not constitute an alternative to powerful canonical quantization. However, its simplicity allows for the incorporation of perturbations, which lead to the "annihilation" and "creation" of solitons, an effect that does not have a classical analog. Examples of this procedure have been presented in [39] in the case of (1+1)-dimensional systems (the KdV [21], mKdV [22], Sawada-Kotera [23] and bi-directional KdV [24, 25] equations) and in [40] for the Kadomtsev-Petviashvili II equation in (1+2) dimensions [26]. In all cases, a Hirota algorithm for the construction of the soliton solutions of the equations [27-31] was exploited.

Construction of soliton solutions of the Sine-Gordon equation via the Hirota algorithm is reviewed in Section 2. Originally developed for the (1+1)-dimensional equation [32], the algorithm generates $N$-soliton solutions of that equation also in (1+2) dimensions, for all $N \geq 1$ [41]. These solutions propagate rigidly at velocities that are lower than the speed of light ($c = 1$). The quantum-mechanical system is described in Section 3. It is based on a Fock space of particles with one coordinate – an angle. A nonlinear functional of the particle number-operators, denoted by U, is constructed. U obeys the Sine-Gordon equation in (1+2) dimensions. All $N$-particle states are eigenstates of U. The eigenvalues are the $N$-soliton solutions of the classical equation. In Section 4, a nonlinear functional of U, which projects the full Fock space onto its multi-particle subspace, is presented. It plays the role of a mass-density operator. Its eigenvalues on multi-particle states are mass densities of localized structures that emulate free, spatially extended, relativistic particles.

The-mass density operator obeys the wave equation, driven by a source term through coupling to Sine-Gordon solitons. Finally, an example for the inclusion of particle interactions, which induce soliton effects that do not have a classical analog, is presented in Section 5.

## 2. Soliton solutions of the Sine-Gordon equation – a review

### 2.1 Solutions in (1+1) dimensions

The (1+1)-dimensional Sine-Gordon equation,

$$\partial_\mu \partial^\mu u + \sin u = 0 \qquad (\mu = 0,1) \ , \tag{1}$$

was shown to be integrable within the framework of the Inverse Scattering formalism [33]. The Hirota algorithm [32] for the construction of its $N$-soliton solutions for all $N \geq 1$ is based on a transformation, which was first presented in the cases of one- and two-soliton solutions in [7,8]:

$$u\left(x;Q^{(N)}\right) = 4\tan^{-1}\left[ g\left(x;Q^{(N)}\right) \big/ f\left(x;Q^{(N)}\right) \right] \ . \tag{2}$$

In Eq. (2),

$$Q^{(N)} \equiv \left\{ q^{(1)}, q^{(2)}, \ldots, q^{(N)} \right\} \ . \tag{3}$$

$x$ and $q$ are, respectively, (1+1)-dimensional coordinate and momentum vectors. In addition:

$$g\left(x;Q^{(N)}\right) = \sum_{\substack{1\leq n\leq N \\ n\ odd}} \left( \sum_{1\leq i_1 < \cdots < i_n \leq N} \left\{ \prod_{j=1}^{n} \varphi\left(x;q^{(i)}\right) \prod_{i_l < i_m} V\left(q^{(i_l)}, q^{(i_m)}\right) \right\} \right) \ , \tag{4}$$

$$f\left(x;Q^{(N)}\right) = 1 + \sum_{\substack{2\leq n\leq N \\ n\ even}} \left( \sum_{1\leq i_1 < \cdots < i_n \leq N} \left\{ \prod_{j=1}^{n} \varphi\left(x;q^{(i)}\right) \prod_{i_l < i_m} V\left(q^{(i_l)}, q^{(i_m)}\right) \right\} \right) \ , \tag{5}$$

$$\varphi\left(x;q^{(i)}\right) = e^{q_\mu^{(i)} x^\mu + \delta^{(i)}} \ , \tag{6}$$

where $\delta^{(i)}$ are constant, arbitrary, phase shifts,

$$q_\mu^{(i)} q^{(i)\mu} = -1 \ , \tag{7}$$

and

$$V\left(q,q'\right)=\frac{1+q_{\mu}q'^{\mu}}{1-q_{\mu}q'^{\mu}} \quad . \tag{8}$$

The solitons show up in the current density:

$$J_{\mu}=\partial_{\mu}u\left(x\right) \quad . \tag{9}$$

Finally, as the solution is constructed in terms of Lorentz scalars ($x\cdot q^{(i)}$, $q^{(i)}\cdot q^{(j)}$, $1 \le i \ne j \le N$), it is also a Lorentz scalar. Namely, for the coordinate vector and the momenta in two reference frames that are connected by a Lorentz transformation, $L$, $u$ obeys:

$$u\left(x,\left\{q^{(1)},...,q^{(N)}\right\}\right)=u\left(\tilde{x},\left\{\tilde{q}^{(1)},...,\tilde{q}^{(N)}\right\}\right) \qquad \left(x=L\cdot\tilde{x}, \qquad q^{(i)}=L\cdot\tilde{q}^{(i)} \quad , \quad 1\le i\le N\right) \quad . \tag{10}$$

(Lorentz invariance of the single-soliton solution in (1+1) dimensions was first pointed out in [8].)

**2.2 Extension to (1+2) dimensions**

Hirota attempted to apply his algorithm to the (1+2)-dimensional version of Eq. (1), namely, with $x$ and $q^{(i)}$ being vectors in the (1+2)-dimensional Minkowski space. This resulted in the construction of one- and two-soliton solutions, but encountered an impasse in the case of three solitons [34]. For a three-soliton solution to exist, one of the three momentum vectors, from which the solution is constructed via Eqs. (2)-(8), had to be a linear combination of the other two. About one decade later on, it was shown that the (1+2)-dimensional Sine-Gordon equation is not integrable under integrability tests employed in nonlinear dynamics [35-38].

The impasse pointed out to by Hirota is the key to the extension of his algorithm to $N$-soliton solutions in both (1+2) and (1+3) dimensions for any $N \ge 3$ [41]. For a solution with $N \ge 3$ solitons to exist, ($N$–2) of the momentum vectors must be linear combinations of two of them. This constraint on the momenta (the eigenvalues in an Inverse Scattering analysis) is the manifestation of the fact that the Sine-Gordon equation in higher space dimensions remains non-integrable. As a result of the constraint, all multi-soliton solutions in (1+3) dimensions can be obtained by applying three-dimensional rotations to (1+2)-dimensional solutions. Therefore, this paper focuses on the (1+2)-dimensional solutions. Their properties are reviewed in the following.

All multi-soliton solutions in (1+2) dimensions, the velocity of which is lower than the speed of light ($c = 1$), are Lorentz transforms of static (time independent) solutions. The latter are constructed through Eqs. (2)-(8) with (1+2)-dimensional momentum vectors of the form:

$$\tilde{q}^{(i)} = \left\{0, \cos\psi_i, \sin\psi_i\right\}, \qquad \left(1 \le i \le N\right) \quad . \tag{11}$$

The two-particle coefficients of Eq. (8) then obtain the form:

$$V\left(\tilde{q}^{(i)}, \tilde{q}^{(j)}\right) = \left(\tan\left(\frac{\psi_i - \psi_j}{2}\right)\right)^2 > 0 \quad . \tag{12}$$

Applying a Lorentz transformation in (1+2) dimensions to the coordinate vector, $x$, and the momentum vectors in a static solution yields a Lorentz invariant solution, which propagates rigidly at the velocity of the transformation.

Finally, as the momentum vectors of Eq. (11) all lie in the $x$-$y$ plane, only two of them are linearly independent. (A careful study of Eqs. (2)-(8) reveals that the choice of the two "basis vectors" is unimportant because in a genuine $N$-soliton solution $q^{(i)} \neq \pm\, q^{(j)}$ for all $1 \le i \neq j \le N$.) This geometrical characteristic of the momentum vectors is retained when they are Lorentz-transformed to a moving frame. Hence, an $N$-soliton solution with $N \ge 2$ depends only on two Lorentz scalars:

$$\xi_i = q^{(i)} \cdot x \qquad \left(i = 1,2\right) \quad . \tag{13}$$

In terms of these two variables, over the space of Hirota-type multi-soliton solutions, the Sine-Gordon equation in (1+2) dimensions is reduced to

$$-\partial_{\xi_1}^2 u + 2\left(q^{(1)} \cdot q^{(2)}\right)\partial_{\xi_1}\partial_{\xi_1} u - \partial_{\xi_2}^2 u + \sin u = 0 \quad . \tag{14}$$

## 2.3 Limiting cases

When any two momentum vectors, $q^{(i)}$, $q^{(j)}$, $1 \le i \ne j \le N$, coincide, the coefficient, $V(q^{(i)}, q^{(j)})$ of Eq. (8), vanishes. The solution then degenerates into one with $N-1$ solitons:

$$\begin{array}{c} g\left(x;Q^{(N)}\right) \Rightarrow g\left(x;Q^{(N-1)}\right) \quad , \quad f\left(x;Q^{(N)}\right) \Rightarrow f\left(x;Q^{(N-1)}\right) \\ u\left(x;Q^{(N)}\right) \Rightarrow u\left(x;Q^{(N-1)}\right) \end{array} \quad . \tag{15}$$

The situation is more complicated when any two vectors become anti-parallel:

$$q^{(i)} \to -q^{(j)} \quad , \quad i \ne j \quad . \tag{16}$$

In this limit, the coefficient $V(q^{(i)}, q^{(j)})$ becomes unbounded (see Eq. (8)). As a result, $g(x,Q^{(N)})$ and $f(x,Q^{(N)})$ (see Eqs. (4) and (5)) become unbounded. However, $u(x,Q^{(N)})$, remains bounded and degenerates into a solution with $N-2$ solitons:

$$\left|g\left(x;Q^{(N)}\right)\right| \to \infty \quad , \quad \left|f\left(x;Q^{(N)}\right)\right| \to \infty \quad , \quad u\left(x;Q^{(N)}\right) \to u\left(x;Q^{(N-2)}\right) \quad . \tag{17}$$

This divergence is avoided if one defines regularized versions of Eqs. (4) and (5):

$$\tilde{g}\left(x;Q^{(N)},\varepsilon\right) = \sum_{\substack{1\le n\le N \\ n\ odd}} \left( \sum_{1\le i_1<\cdots<i_n\le N} \left\{ \prod_{j=1}^{n} \varphi\left(x;q^{(i)}\right) \prod_{i_l<i_m} \frac{V\left(q^{(i_l)},q^{(i_m)}\right)}{\sqrt{1+\left(\varepsilon V\left(q^{(i_l)},q^{(i_m)}\right)\right)^2}} \right\} \right) \quad , \tag{18}$$

$$\tilde{f}\left(x;Q^{(N)},\varepsilon\right) = 1 + \sum_{\substack{2\le n\le N \\ n\ even}} \left( \sum_{1\le i_1<\cdots<i_n\le N} \left\{ \prod_{j=1}^{n} \varphi\left(x;q^{(i)}\right) \prod_{i_l<i_m} \frac{V\left(q^{(i_l)},q^{(i_m)}\right)}{\sqrt{1+\left(\varepsilon V\left(q^{(i_l)},q^{(i_m)}\right)\right)^2}} \right\} \right) \quad , \tag{19}$$

$$\tilde{u}\left(x;Q^{(N)},\varepsilon\right) = 4\tan^{-1}\left[\tilde{g}\left(x;Q^{(N)},\varepsilon\right) \big/ \tilde{f}\left(x;Q^{(N)},\varepsilon\right)\right] \quad . \tag{20}$$

In all cases, except for that of Eq. (16), the limit $\varepsilon \to 0$ yields the results of Eqs. (2)-(8). In the case of Eq. (16), one first applies Eq. (16), and then considers the limit of small $\varepsilon$:

$$\tilde{g}\left(x;Q^{(N)},\varepsilon\right)\to\frac{1}{\varepsilon}g\left(x;Q^{(N-2)}\right)+O(1)\quad,\quad\tilde{f}\left(x;Q^{(N)},\varepsilon\right)\to\frac{1}{\varepsilon}f\left(x;Q^{(N-2)}\right)+O(1)$$
$$\tilde{u}\left(x;Q^{(N)},\varepsilon\right)\to u\left(x;Q^{(N-2)}\right)+O(\varepsilon) \quad . \tag{21}$$

This regularization procedure is not required in the classical case, but turns out to be useful in the construction of the quantum-mechanical system, described in the following.

## 3. Quantum-mechanical system

### 3.1 Fock space

Consider a Fock space of Bosons (Fermions would do just as well), with one coordinate, an angle, $0 \le \psi \le 2\pi$. The particle-number operators are given in terms of creation and annihilation operators:

$$\mathrm{N}_{\psi}=a^{\dagger}_{\psi}\,a_{\psi}\qquad\qquad\left(\left[a_{\psi},a^{\dagger}_{\psi'}\right]=\delta\left(\psi-\psi'\right)\right)\ . \tag{22}$$

An $N$-particle state with all angles different will be denoted by

$$\left|\Psi_N\right\rangle=\left(\prod_{i=1}^{N}a^{\dagger}_{\psi_i}\right)\left|0\right\rangle\ . \tag{23}$$

### 3.2 Sine-Gordon operators – static frame

The first step is the definition of operator versions of Eqs. (4) and (5) in the static frame, where Eqs. (11) and (12) hold. This requires some care, since $V\left(\tilde{q}^{(l)},\tilde{q}^{(m)}\right)$ become unbounded when $\left|\psi_l-\psi_m\right|\to\pi$ for any $1 \le l \ne m \le N$ (see Eq. (12)). To this end, one first defines operator versions of the regularized entities of Eqs. (18) and (19):

$$\tilde{\mathrm{G}}(x;\varepsilon)=\sum_{\substack{n\ge1\\ n\ odd}}^{\infty}\frac{1}{n!}\int_0^{2\pi}\int_0^{2\pi}\ldots\int_0^{2\pi}\left(\prod_{i=1}^{n}\varphi\left(x;\tilde{q}^{(i)}\right)\mathrm{N}_{\psi_i}\right)\left(\prod_{1\le l<m\le n}\left(\frac{V\left(\tilde{q}^{(l)},\tilde{q}^{(m)}\right)}{\sqrt{1+\left(\varepsilon V\left(\tilde{q}^{(l)},\tilde{q}^{(m)}\right)\right)^2}}\right)\right)d\psi_1\,d\psi_2\cdots d\psi_n\quad , \tag{24}$$

$$\tilde{\mathrm{F}}(x;\varepsilon)=1+\sum_{\substack{n\ge2\\ n\ even}}^{\infty}\frac{1}{n!}\int_0^{2\pi}\int_0^{2\pi}\ldots\int_0^{2\pi}\left(\prod_{i=1}^{n}\varphi\left(x;\tilde{q}^{(i)}\right)\mathrm{N}_{\psi_i}\right)\left(\prod_{1\le l<m\le n}\left(\frac{V\left(\tilde{q}^{(l)},\tilde{q}^{(m)}\right)}{\sqrt{1+\left(\varepsilon V\left(\tilde{q}^{(l)},\tilde{q}^{(m)}\right)\right)^2}}\right)\right)d\psi_1\,d\psi_2\cdots d\psi_n\ . \tag{25}$$

Inclusion of the denominators with the auxiliary parameter, $\varepsilon$, in Eqs. (24) and (25) ensures convergence of the integrals. For fixed $\varepsilon$, the matrix elements of $\tilde{\mathrm{G}}(x;\varepsilon)$ and $\tilde{\mathrm{F}}(x;\varepsilon)$ are finite also in the limit, in which any two momentum vectors become anti-parallel (see Eq. (16)).

$\tilde{\mathrm{G}}(x;\varepsilon)$ and $\tilde{\mathrm{F}}(x;\varepsilon)$ are diagonal operators. $N$-particle states in the Fock space are their eigenstates; the eigenvalues are, respectively, $\tilde{g}\left(x;Q^{(N)},\varepsilon\right)$ and $\tilde{f}\left(x;Q^{(N)},\varepsilon\right)$ of Eqs. (18) and (19):

$$\begin{aligned} \tilde{\mathrm{G}}(x;\varepsilon)\left|\Psi_N\right\rangle &= \tilde{g}\left(x;Q^{(N)},\varepsilon\right)\left|\Psi_N\right\rangle \\ \tilde{\mathrm{F}}(x;\varepsilon)\left|\Psi_N\right\rangle &= \tilde{f}\left(x;Q^{(N)},\varepsilon\right)\left|\Psi_N\right\rangle \end{aligned} \quad . \tag{26}$$

In Eqs. (26), $Q^{((N)}$ is the set of $N$ momentum vectors defined in Eq. (11), for which the $N$-soliton solution is static:

$$Q^{(N)} = \left\{\tilde{q}^{(1)},...,\tilde{q}^{(N)}\right\} \ . \tag{27}$$

Noting that $\tilde{\mathrm{F}}(x;\varepsilon)$ is positive definite and commutes with $\tilde{\mathrm{G}}(x;\varepsilon)$, one defines an operator analog of Eq. (20):

$$\tilde{\mathrm{G}}(x;\varepsilon)\tilde{\mathrm{F}}(x;\varepsilon)^{-1} = \tan\left[\frac{1}{4}\tilde{\mathrm{U}}(x;\varepsilon)\right] \ . \tag{28}$$

$\tilde{\mathrm{U}}(x;\varepsilon)$ is a diagonal operator. All $N$-particle states are its eigenstates. The eigenvalues are $\tilde{u}\left(x;Q^{(N)},\varepsilon\right)$ of Eq. (20):

$$\tilde{\mathrm{U}}(x;\varepsilon)\left|\Psi_N\right\rangle = \tilde{u}\left(x;Q^{(N)},\varepsilon\right)\left|\Psi_N\right\rangle \ . \tag{29}$$

One now defines a diagonal operator, U($x$), through its matrix elements:

$$\left\langle\Psi_N\right|\mathrm{U}(x)\left|\Psi_N\right\rangle = \lim_{\varepsilon\to 0}\left\langle\Psi_N\right|\tilde{\mathrm{U}}(x;\varepsilon)\left|\Psi_N\right\rangle = \lim_{\varepsilon\to 0}\tilde{u}\left(x;Q^{(N)},\varepsilon\right) = u\left(x;Q^{(N)}\right) \ . \tag{30}$$

Eq. (30) is valid also if any angle, $\psi_i$, in the state, $\left|\Psi_N\right\rangle$, is populated by $n_i > 1$ particles. The only change is the addition of a constant term, log $n_i$, to the free constant phase, $\delta^{(i)}$, in Eq. (6).

Thanks to Eqs. (2)-(8), the static soliton solutions of Eq. (1) in (1+2) dimensions are the eigenvalues of U($x$). As a result, U($x$) obeys the Sine-Gordon equation in the static frame:

$$-\partial^2_{x_1}\mathrm{U} - \partial^2_{x_2}\mathrm{U} + \sin\mathrm{U} = 0 \quad . \tag{31}$$

### 3.3 Moving frame

The eigenvalues of the operator, U($x$), defined in Eq. (30), are Lorentz scalars. Hence, this operator describes the dynamics also in any moving frame of reference. When acting on the particle states of Eq. (23), its eigenvalues are $u(x;Q^{(N)})$ in any moving frame that is obtained from the rest frame by a Lorentz transformation, $L$. The reason is that the same applies to the regularized operators, $\tilde{\mathrm{G}}(x;\varepsilon)$ and $\tilde{\mathrm{F}}(x;\varepsilon)$. For example, denoting the coordinate- and momentum-vectors in the rest frame, respectively, by $\tilde{x}$ and $\tilde{q}^{(i)}$, then, in a moving reference frame, $\tilde{\mathrm{G}}(x;\varepsilon)$ obeys:

$$\tilde{\mathrm{G}}(x;\varepsilon) = \tilde{\mathrm{G}}(\tilde{x};\varepsilon) \qquad \left(x = L\cdot\tilde{x}, \qquad q^{(i)} = L\cdot\tilde{q}^{(i)} \quad , \quad 1 \le i \le N\right) \quad . \tag{32}$$

As a result, in a moving frame, U($x$) obeys the (1+2)-dimensional Sine-Gordon equation

$$\partial_\mu\partial^\mu\mathrm{U} + \sin\mathrm{U} = 0 \quad . \tag{33}$$

Consequently, replacing $u$ by U in the classical Lagrangian that generates the Sine-Gordon equation, the latter becomes the Lagrangian for the quantum system:

$$L = \int\left\{\tfrac{1}{2}\partial_\mu u\,\partial^\mu u - (1-\cos u)\right\}d^2\vec{x} \Rightarrow \int\left\{\tfrac{1}{2}\partial_\mu\mathrm{U}\,\partial^\mu\mathrm{U} - (1-\cos\mathrm{U})\right\}d^2\vec{x} \quad . \tag{34}$$

## 4. Spatially extended relativistic "particles"

### 4.1 Projection operator as mass generator

Consider the operator,

$$R[\mathrm{U}] = \tfrac{1}{2}\partial_\mu\mathrm{U}\,\partial^\mu\mathrm{U} + (1-\cos\mathrm{U}) \quad . \tag{35}$$

The $N$-particle states are eigenstates of $R$[U]. The eigenvalues are equal to $R[u]$, with $u$ – an $N$-soliton solution. The properties of $R[u]$ have been discussed in [42] and are summarized in the following. On a single-particle state, $u$ is a single-soliton solution, for which $R[u]$ vanishes. Hence, $R$[U] is a projection operator, vanishing on the single-particle subspace. On a multi-particle state, the eigenvalue, $R[u]$, is a structure, which is localized around soliton junctions. It preserves its shape in time as it moves together with the soliton solution, $u$, at the constant velocity of the latter.

If one defines the "mass" of the structure in a moving frame as

$$m = \int R[u] dx\, dy \quad , \tag{36}$$

then the structure emulates a free, spatially extended, massive relativistic particle. The reason is that $R[u]$ is a Lorentz scalar. As a result, the mass in a frame that moves at velocity $v$ differs from the rest mass, $m_0$ (the mas of $R[u]$ - computed on a static multi-solution solution), by the Jacobian of the space part of the Lorentz transformation, leading to:

$$m = m_0 \Big/ \sqrt{1 - v^2} \quad . \tag{37}$$

Finally, such a spatially extended "particle" is, in fact, a composite object; it is the eigenvalue of an operator over a state that contains several particles.

### 4.2 Equation of motion of mass generating operator

The static solution (denoted in the following by $u^{(S)}$) obeys the time-independent Sine-Gordon equation in two space dimensions:

$$-\partial_x^{\,2} u^{(S)} - \partial_y^{\,2} u^{(S)} + \sin u^{(S)} = 0\,. \tag{38}$$

Repeated application of Eq. (38) yields that $R[u^{(S)}]$ obeys the following equation:

$$
\begin{aligned}
&-\partial_x{}^2 R\left[u^{(S)}\right]-\partial_y{}^2 R\left[u^{(S)}\right]= \\
&2\left(\left(u^{(S)}{}_{xy}\right)^2-u^{(S)}{}_{xx}\,u^{(S)}{}_{yy}\right)= \\
&2\left(1-\left(\tilde{q}^{(1)}\cdot\tilde{q}^{(2)}\right)^2\right)\left(\left(u^{(S)}_{\xi_1\xi_2}\right)^2-u^{(S)}_{\xi_1\xi_1}\,u^{(S)}_{\xi_2\xi_2}\right)^2
\end{aligned}
\quad . \tag{39}
$$

In Eq. (39), the entity expressed in terms of the Lorentz scalar variables of Eq. (13) is the Lorentz-invariant form of the source term. As $R$[U] is diagonal, Eq. (39) leads to the conclusion that, in the rest frame, $R$[U] obeys the equation:

$$
-\partial_x{}^2 R[\mathrm{U}]-\partial_y{}^2 R[\mathrm{U}]=2\left(\left(\mathrm{U}_{xy}\right)^2-\mathrm{U}_{xx}\,\mathrm{U}_{yy}\right) \quad . \tag{40}
$$

Thus, in the rest frame, the mass-generating operator obeys a (time-independent) wave equation, driven by a source term that is constructed from the soliton-generating operator, U. Applying the source term on the r.h.s. of (40) to $N$-particle states, one finds its eigenvalues:

$$
\begin{aligned}
&2\left(\left(\mathrm{U}_{xy}\right)^2-\mathrm{U}_{xx}\,\mathrm{U}_{yy}\right)\left|\Psi_N\right\rangle= \\
&2\left(\left(u^{(S)}_{xy}\right)^2-u^{(S)}_{xx}\,u^{(S)}_{yy}\right)\left|\Psi_N\right\rangle= \\
&2\left(1-\left(\tilde{q}^{(1)}\cdot\tilde{q}^{(2)}\right)^2\right)\left(\left(u^{(S)}_{\xi_1\xi_2}\right)^2-u^{(S)}_{\xi_1\xi_1}\,u^{(S)}_{\xi_2\xi_2}\right)^2\left|\Psi_N\right\rangle
\end{aligned}
\quad . \tag{41}
$$

One first observes that, as expected, the source term vanishes when $N = 1$ (one particle state, corresponding to a single soliton). This is easiest seen through the Lorentz invariant form of the eigenvalue, by noting that, in this limit, all momentum vectors coincide, and the coefficient $\left(1-\left(\tilde{q}^{(1)}\cdot\tilde{q}^{(2)}\right)^2\right)$ vanishes. Next, one observes that transforming Eq. (40) to a moving frame by a Lorentz transformation, denoted by $L$, it becomes the relativistic wave equation driven by a Lorentz invariant driving term:

$$
\partial_\mu\partial^\mu R[\mathrm{U}]=2L\cdot\left\{\left(\left(\mathrm{U}_{xy}\right)^2-\mathrm{U}_{xx}\,\mathrm{U}_{yy}\right)\right\}^{(S)} \quad . \tag{42}
$$

In Eq. (42), the superscript $S$ signifies that the entity in curly brackets is constructed in the static frame, and then transformed by the Lorentz transformation to a moving frame. The transformed form of the source term on the r.h.s. of Eq. (42) is complicated and not instructive. However, its eigenvalues on states in the Fock space retain their relativistic invariant form given in Eq. (41). (This last statement is confirmed directly by repeated exploitation of Eq. (1) (in (1+2) dimensions) in the calculation of $\partial_\mu \partial^\mu R[u]$, the eigenvalue of $\partial_\mu \partial^\mu R[\mathrm{U}]$ when applied to an $N$-particle state.)

**5. Quantum-mechanical effects that do not have a classical analogue**

When a small perturbation is added to the classical equation, Eq. (1), most often, one can obtain approximations to the solution of the perturbed equation only through a divergent asymptotic perturbation series. As the higher-order corrections are functionals of the zero-order term (an $N$-soliton solution), they do not have the capacity to change the identity of the zero-order solitons through any finite order of the expansion. This shortcoming of the classical case can be overcome in the quantum-dynamical system. One can incorporate perturbations that contain Fock-space-particle interactions, which have the capacity to "create" and "annihilate" solitons in any order of the perturbative expansion of the solution. As an example, consider the following perturbed version of Eq. (1) in (1+2) dimensions:

$$\partial_\mu \partial^\mu u + \sin u + \varepsilon\left(u\cos u - \sin u\right) = 0 \quad . \tag{43}$$

The perturbation has been chosen so as to simplify the analysis. Specifically, through $O(\varepsilon)$, the approximate solution is given by:

$$u = \left(1-\varepsilon\right)u_0 \quad . \tag{44}$$

The zero-order term, $u_0$, coincides with the soliton solution of the (1+2) dimensional Sine-Gordon equation, constructed via Eqs. (2)-(8).

A possible quantized version of Eq. (43) is:

$$\partial_\mu \partial^\mu \mathrm{U} + \sin \mathrm{U} + \varepsilon \mathrm{P}\left(\mathrm{U}\cos\mathrm{U} - \sin\mathrm{U}\right) = 0 \quad . \tag{45}$$

In Eq. (45), choose the operator, P, to have the capacity to create and annihilate particles:

$$\mathrm{P}=\int_0^{2\pi}\int_0^{2\pi} c(\psi,\psi')a^\dagger_\psi a_{\psi'}\,d\psi\,d\psi' \qquad \left(\int_{-\infty}^{+\infty}\int_{-\infty}^{+\infty}\left|c(\psi,\psi')\right|^2 d\psi\,d\psi' < \infty\right) \quad . \tag{46}$$

Let us now expand U through $O(\varepsilon)$:

$$\mathrm{U}=\mathrm{U}_0+\varepsilon\,\mathrm{U}_1 \quad . \tag{47}$$

The zero-order term, $\mathrm{U}_0$, follows the result of the classical case. It is the operator that is defined by Eq. (30).

Substituting Eq. (47) in Eq. (45), exploiting Eq. (30) for $\mathrm{U}_0$, and noting that $\mathrm{U}_0$ and $\mathrm{U}_1$ need not commute, the $O(\varepsilon)$ part of Eq. (45) is found to be:

$$\partial_\mu\partial^\mu \mathrm{U}_1+\sum_{k=0}^{\infty}\frac{(-1)^k}{(2k+1)!}\sum_{j=1}^{2k+1}U_0^{\,j-1}U_1U_0^{\,2k+1-j}+\mathrm{P}\left(\mathrm{U}_0\cos\mathrm{U}_0 - \sin\mathrm{U}_0\right)=0 \quad . \tag{48}$$

As $\mathrm{U}_0$ is diagonal, the matrix elements of Eq. (48) between two states are given by:

$$\begin{aligned}&\partial_\mu\partial^\mu\langle\vec{\psi}|\mathrm{U}_1|\vec{\psi}'\rangle+\langle\vec{\psi}|\mathrm{U}_1|\vec{\psi}'\rangle\frac{\sin\left[u_0(\vec{\psi}')\right]-\sin\left[u_0(\vec{\psi})\right]}{u_0(\vec{\psi}')-u_0(\vec{\psi})}\\ &+\langle\vec{\psi}|\mathrm{P}|\vec{\psi}'\rangle\left(u_0(\vec{\psi}')\cos u_0(\vec{\psi}') - \sin u_0(\vec{\psi}')\right)=0\end{aligned} \quad . \tag{49}$$

In Eq. (49), $u_0(\vec{\psi})$ and $u_0(\vec{\psi}')$ are zero-order terms, as in in Eq. (44). They are solutions of the Sine-Gordon equation (1+2) dimensions, constructed through Eqs. (2)-(8), with momentum vectors obtained by a Lorentz transformation of the vectors in the static frame, given by Eq. (11).

The states $|\vec{\psi}\rangle$ and $|\vec{\psi}'\rangle$ may differ by the numbers of particles and/or the angle-coordinates of particles. However, based on Eq. (46), the matrix elements of P vanish when the numbers of particles in the two states are different, or, when they are equal, if the states differ by more than one angle-coordinate. Hence, such off-diagonal matrix elements of $\mathrm{U}_1$ are solutions of the homogeneous version of Eq. (49), namely, without the driving term that contains P. Such matrix elements do not reflect the effect of the perturbation, and can be avoided by appropriate initial and boundary conditions. The perturbation contributes physically interesting effects only when the numbers of particles in $|\vec{\psi}\rangle$ and $|\vec{\psi}'\rangle$ coincide, and if the two states differ by, at most one angle:

$$\vec{\psi}=\left\{\psi^{(1)},...,\psi^{(k-1)},\psi^{(k)},\psi^{(k+1)},...,\psi^{(N)}\right\} \quad , \quad \vec{\psi}'=\left\{\psi^{(1)},...,\psi^{(k-1)},\psi'^{(k)},\psi^{(k+1)},...,\psi^{(N)}\right\} \quad . \tag{50}$$

In Eq. (50), $1 \le k \le N$, and $\psi^{(k)}$ and $\psi'^{(k)}$ may be equal or different.

The diagonal matrix elements in Eq. (49) obey:

$$\partial_\mu \partial^\mu \langle\vec{\psi}|\mathrm{U}_1|\vec{\psi}\rangle + \langle\vec{\psi}|\mathrm{U}_1|\vec{\psi}\rangle \cos\left[u_0(\vec{\psi})\right] + \langle\vec{\psi}|\mathrm{P}|\vec{\psi}\rangle\left(u_0(\vec{\psi})\cos u_0(\vec{\psi}) - \sin u_0(\vec{\psi})\right) = 0 \quad . \tag{51}$$

The solution of Eq. (51) follows the pattern of the $O(\varepsilon)$ correction in the classical case, Eq. (44):

$$\langle\vec{\psi}|\mathrm{U}_1|\vec{\psi}\rangle = -\langle\vec{\psi}|\mathrm{P}|\vec{\psi}\rangle u_0(\vec{\psi}) = -\left(\sum_{i=1}^{N} c\left(\psi^{(i)},\psi^{(i)}\right)\right) u_0(\vec{\psi}) \quad . \tag{52}$$

However, off-diagonal matrix elements will mix in other solitons (i.e. the momentum vector of one of the solitons is changed). Thus, already in $O(\varepsilon)$, the solution, U, will contain corrections, in which one soliton is "annihilated" and another one is "created" – something that is impossible in an order-by-order expansion of the solution in the classical case, Eq. (43).

**6. Concluding comments**

The quantum-mechanical system constructed above offers interesting possibilities.

1) It allows for the incorporation of Fock-space particle interactions, which have the capacity to "create" or "annihilate" solitons - an effect that does not have a classical analog.

2) The structures that emulate free, spatially extended, massive relativistic particles are *not* wave functions of bound states of a quantum mechanical system. Rather, they are eigenvalues of a projection operator. Mass is generated though a soliton-based source term, which drives the wave equation obeyed by this operator. Finally, these structures are composite entities, connected with states of two, or more, particles in the Fock space.

References

1. Frenkel Y. and Kontorova T., *J. Phys. (USSR)* **1**, 137-149 (1939).
2. A. Seeger, H. Donth and A. Kochendorfer, *Z. Physik* **134**, 173- (1953).
3. J.K. Perring and T.H. Skyrme, *Nucl. Phys.* **31**, 550-555 (1962).
4. B.D. Josephson, *Adv. Phys.* **14**, 419-451 (1965).
5. A.C. Scot, *Amer. J. Phys*. **37**,52-61 (1969).
6. J. Rubinstein, *J. Math. Phys.* **11** 258-266 (1970).
7. G.L. Lamb Jr., *Rev. Mod. Phys.* **43**, 99-124 (1971).
8. A. Barone, F. Esposito, C.J. Magee and A.C. Scott, *Rev. Nuov. Cim.* **1**, 227-267 ((1971).
9. A.C. Scott, F.Y.F. Chu and D.W. McLaughlin, *Proc. IEEE* **61**,1443-1483 (1973).
10. S. Coleman, *Phys. Rev. D* **11**, 2068- (1975).
11. S. Mandelstam, *Phys. Rev. D* **11** 3026- 3030 (1975).
12. R.F. Dashen, B. Hasslacher and A. Neveu, *Phys, Rev. D* **11**, 3424- (1975).
13. C.G. Callan and D.J. Gross, *Nucl. Phys*. B **93**, 29-55 (1975).
14. S. Coleman, *Phys. Rev*. D **11**, 2088-2097 (1975).
15. L.D. Fadeev, V.E. Korepin *Phys. Rep.* **420**, 1-78 (1978).
16. E. Witten, *Nucl. Phys*. B **160**, 57-115 (1979).
17. S.N.M. Ruijsenaars, *Commun. Math. Phys.* **110**, 191-213 (1987).
18. O. Babelon, D. Berbard and F.A. Smironov, *Commun. Math. Phys.* **182**, 319-354 (1996).
19. S.N.M. Ruijsenaars, *NATO Sci. Ser. II: Math. Phys. & Chem.* **35**, 273-292 (2001).
20. A. Mikhailov, *J. Geom. Phys.* **56**, 2429-2445 (2006).
21. D.J. Korteweg and G. De Vries, *Phil. Mag.* **39**, 422-443 (1895).
22. R.M. Miura, *J. Math. Phys.* **9**, 1202-1204 (1968).
23. K. Sawada and T. Kotera, *Progr. Th. Phys.* **51**, 1355-1367 (1974).
24. R. Hirota, *J. Math. Phys.* **14**, 810-815 (1973).
25. P.J. Caudrey, R.K. Dodd and J.D. Gibbon, *Proc. Roy. Soc. London A* **351,** 407-422 (1976).
26. B.B. Kadomtsev and V.I. Petviashvili, *Sov. Phys. Dokl*. **15**, 539–541 (1970).
27. R. Hirota, *Phys. Rev. Lett.* **27**, 1192-1194 (1971).
28. R. Hirota, *J Phys. Soc. Japan* **33**, 1456-1458 (1972).
29. R. Hirota, *J. Math. Phys.* **14** (1973) 810-815.
30. R. Hirota, *J. Phys. Soc. Japan*, **55** (1986) 2137-2150.
31. R. Hirota, Y. Ohta and J. Satsuma, *Progr. Th. Phys. Suppl*. **94** (1988) 59-72.
32. R. Hirota, *J. Phys. Soc. Japan* **33**, 1459-1463 (1972).
33. M.J. Ablowitz, *Phys. Rev. Lett.*, **30**, 1262-1264 (1973).
34. R. Hirota, J. Phys. Soc. Japan, **35**, 1566 (1973).
35. M.J. Ablowitz, A. Ramani and H. Segur, *J. Math. Phys.*, **21**, 715-721 (1980).
36. J. Weiss, M. Tabor and G. Carnevale, *J. Math. Phys.*, **24**, 522-526 (1983)
37. J. Weiss, *J. Math. Phys.*, **25**, 2226-2235 (1984).
38. P.A. Clarkson, *Lett. Math. Phys.*, **10**, 297-299 (1985).
39. Y. Zarmi, *Phys. Rev. E* **83**, 056606 (2011).
40. Y. Zarmi, *J. Math. Phys.* **54**, 063515 (2013).
41. Y. Zarmi, *J. Math. Phys.* **54**, 013512 (2013).
42. Y. Zarmi, arXiv:1310.4044 [nlin.SI].